# MBE grown preferentially oriented CdMgO alloy on *m*- and *c*-plane sapphire substrates

A. Adhikari, A. Lysak, A. Wierzbicka, P. Sybilski, A. Reszka, B. S. Witkowski, E. Przezdziecka

*Institute of Physics, Polish Academy of Sciences, Al. Lotników 32/46, 02-668 Warsaw, Poland*

## Abstract

Unlike other II-VI semiconductors, CdO-based transparent oxide has great potential application for the fabrication of many optoelectronic devices. In this work, we study the growth of $Cd_xMg_{1-x}O$ alloys on *m*- and on *c*-plane sapphire substrates in Cd-rich to Mg-rich conditions using the plasma-assisted molecular beam epitaxy method. A structural and morphological study of CdMgO random alloys was carried out using X-ray diffraction and Atomic Force Microscope (AFM) techniques whereas composition analysis was done by Energy-dispersive X-ray (EDX) spectroscopy method. The optical properties of thin films were investigated by UV-Vis spectroscopy at room temperature. X-ray analysis confirmed the presence of cubic rock salt structure with <111> CdMgO crystallographic orientation on *c*-plane sapphire and <110> CdMgO preferential orientation on *m*-plane sapphire. The surface roughness was measured by the AFM. From the absorption curve, the optical bandgaps were determined using Tauc relation and it was found that the bandgap of films is influenced by the incorporation of $Mg^{2+}$ ions into the CdO lattice. Bowing parameter was calculated both for samples on *m*- and *c*- sapphires.

Keywords: molecular beam epitaxy, CdO, MgO, ternary alloys, bowing parameter

## 1. Introduction

II-VI oxides and their alloys have recently emerged as a strategic material system for designing many technological photovoltaic applications[1,2]. Their physical stability, high transmittance, and low resistivity features make it an alternative to the III-nitrides system which is widely used in transparent electronics. The recent surge in the manufacturing of fully transparent electronics has sparked significant research interest in finding new, high-performance transparent conducting oxides (TCOs) materials to meet this growing demand. Advancements in the growing technology of group-II binary oxides of Zn, Cd, Mg, and related alloys create opportunities for their use in a wide range of device application purposes. The unique combination of broad bandgap tunability and high exciton binding energy in wurtzite and cubic MgO-ZnO-CdO compounds can benefit a number of modern device applications operating in a wide range of wavelengths[3].

CdO is an n-type semiconductor that usually exists in a cubic rocksalt (RS) crystal structure (space group; $F\bar{m}3m$). Having features of high electron concentration ($\sim 10^{21}/cm^3$), high electron mobility ($\sim 180\ cm^2V^{-1}s^{-1}$), low electron effective mass ($0.21m_o$), high transparency, low electrical resistivity ($\sim 10^{-4}\ \Omega cm$), and high static permittivity ($\varepsilon_o=21.9$)[4,5] makes it one of the oldest TCO for solar cell photovoltaic application. Apart from that CdO has various range of applications such as phototransistors, liquid crystal displays, photodiodes, IR detectors, and anti-reflection coatings[6,7]. However, despite of such advantages, due to a relatively low optical bandgap it is not used widely in short-wavelength optoelectronic application purposes.

Theoretical calculations suggested along with a direct bandgap of 2.3 eV, CdO has two low energy indirect band gaps of 1.2 eV ($\Sigma_3$ and $\Gamma_1$) and 0.8 eV ($L_3$ and $\Gamma_1$), which were agreed with experimental results[8–10]. It is previously reported that CdO can be doped with Boron[11], Aluminium[12], Indium[13], Copper[14], Nickel[15], Fluorine[16], Samarium[17] by various growth techniques in order to widen its optical bandgap few eVs. Another interesting effect of CdO is that bandgap engineering, in which CdO can be alloyed with ZnO ($E_g$=3.37 eV) and MgO ($E_g$= 7.8 eV) binary oxides to tune the bandgap from 2.3 eV to 7 eV and form various thermodynamically stable alloys, hetero-and quantum well structures[18,19]. Homogeneous alloys without any crystal phase segregation can be obtained of CdO and MgO as both exist in the same crystal structure (rocksalt RS cubic structure) whereas, it is a problem for ZnCdO and ZnMgO alloys because of the coexistence of wurtzite and rocksalt phases in a middle composition range[18,20].

Previously Chen *et al.* [21] synthesized CdMgO thin film on the glass substrate (up to 28% Mg concentration) by radio frequency magnetron sputtering technique and found out that incorporation of Mg into CdO lattice results in modification of the electronic structures of the alloys. Guia *et al.* [22] studied Mg-rich CdMgO alloy on *r*-plane sapphire substrate prepared by metal-organic chemical vapour deposition (MOCVD) technique and observed phase separation at Cd concentration above 27%. They also found that the growth temperature plays an important role in preparing Mg-rich to Cd-rich CdMgO random alloys[23]. Lee *et al.* [24] reported In doped CdMgO alloys with 40% Mg concentration used for UV-transparent electron emitter for thin-film photovoltaics technologies. Theoretical investigations on CdMgO alloys have provided further understanding of the structural and optical properties[25,26]. Recently Bashir *et al.* [27] reported the annealing effect of CdMgO nanocomposite deposited by co-evaporation technique on the glass substrate. Recently, superlattice structures (SLs) of CdO and MgO layers grown by plasma-assisted MBE technique on *r*-plane sapphire substrate were reported by Przezdziecka *et al.*[28]. They have observed that the optical bandgap of the SL structures can be tuned from 2.6 to 6 eV by varying the thickness of CdO from 1 to 12 monolayers while keeping the MgO layer thickness constant at 4 monolayers[29].

Despite of possibilities to tune the bandgap in a wide wavelength range, there have been few experimental reports on CdMgO random alloys in a stable phase. In this work, we have incorporated Mg ions into CdO lattice and synthesized CdMgO random alloy structures by plasma-assisted molecular beam epitaxy (PA-MBE) technique on *m*- and *c*- plane sapphire substrates. MBE growth technique will allow us to produce highly crystalline layers with low densities of defects and impurities which are typically required for optimum semiconductor device performance. To provide a more robust understanding of Mg incorporation into CdO lattice, we have decided to grow both Cd-rich to Mg-rich CdMgO random alloys controlled by the growth parameters. The range of growth conditions within which both phases coexist has been analyzed, gives us new insights on the flexibility of incorporating the foreign Mg ions into the host CdO matrix of this alloy.

## 2. Experimental technique

CdMgO random alloys were grown by PA-MBE technique on *m*- and *c*- plane sapphire substrate using a Riber Compact 21B system equipped with conventional Knudsen effusion cells of pure Mg (6N), Cd (6N), and radiofrequency plasma cell for oxygen source. Prior to the growth process, the substrates were chemically etched with $H_2O_2$ and $H_2SO_4$ (1:1 ratio) in order to remove organic and metallic contaminations of the substrates. Following the chemical etching process, the substrates were annealed in the load chamber at 150°C for an hour and finally oxidized for 30 minutes in the growth chamber at 700°C with a flow of oxygen. In order to get Cd-rich to Mg-rich conditions, the Cd and Mg fluxes were controlled by varying Cd and Mg effusion cell temperature, respectively. The growth temperature of the samples was measured by a thermocouple placed behind the substrate. Under the employed growth conditions, the samples were grown at 360°C for 4 hours with oxygen flow 2.5 ml/min, and the rf power of plasma was fixed at 384 W.

Structural investigations were performed by XRD measurement using a PANalytical X'Pert Pro MRD diffractometer equipped with Cu anode X-ray source, 2-bounce Ge (200) hybrid monochromator, and Ge analyzer located in front of a Pixel detector. $Cu_{k\alpha 1}$ radiation of wavelength 1.5406Å was used for all the experiments. More details you can find elsewhere [30]. The compositional study of Cd and Mg concentration in CdMgO alloys was carried out using Hitachi SU-70 scanning electron microscope equipped with Thermo Scientific energy-dispersive X-ray (EDX) spectrometer with silicon drift X-ray detector and Noran System 7. The spectra and elemental distribution maps were acquired with an accelerating voltage of 6 kV. Film thickness were measured using SEM image, which resulted in alloys film thickness at about 300-380 nm. The optical absorption measurements of these random alloys were performed in the 200-700 nm wavelength range at room temperature using a Carry 5000 UV-Vis/NIR spectrophotometer from Agilent Technologies equipped with a PbS detector. The surface morphology was examined by Atomic Force Microscope (Bruker Dimension Icon) using tapping mode.

## 3. Results and Discussion

### 3.1. Structural Properties

The structural analysis of CdMgO random alloys on *m*- and *c*- plane sapphire substrate was done using XRD measurement over a range of 2θ from 30° to 100°. **Fig. 1**a and **Fig. 2**a show the full XRD scans of CdMgO alloys on *m*-plane and *c*-plane sapphire substrates respectively. The substrates diffraction peaks of *m*-, and *c*-plane sapphire are clearly distinguishable[23]. For pure CdO on *m*-plane sapphire, the Bragg peak at 55.3° which is assigned to 220 diffraction peak is clearly visible. In case of pure CdO on *c*-plane sapphire, the Bragg peaks at 33.08°, 38.25°, 64.48°, 69.19°, 82.14° corresponding to 111, 200, 311, 222, 400 diffraction peaks, are detected. These reflections are identified to originate from the cubic RS structure of CdO (JCPDF card no. 652908). So in the case of *m*-plane sapphire, only one orientation is observed, whereas in the case of the *c*-plane substrate the polycrystalline character with strong preferential <111> orientation is detected. Preferential <111> orientation of CdO and MgO as well as obtained by MOCVD CdMgO layers[22] was previously reported on *c*-plane oriented sapphire. The information about the CdMgO growth on *m*-plane sapphire is unavailable in literature but pure CdO grown on *m*-plane sapphire reviled <110> preferential orientation[31].

In case of growth on the *m*-plane sapphire substrate, the peak shift of 220 CdMgO (2θ~55°) with the increasing of Mg concentration to higher angles is observed. In case of growth on *c*-plane sapphire, 111 CdMgO peak shift (near 2θ close to 33°) is noticed. The shifting of peaks corresponding to (110) and (111) diffraction planes are shown in **Fig. 1**b and **Fig. 2**b, respectively. The shift of diffraction peak towards a higher angle is correlated with the progressive incorporation of Mg in CdO. From the XRD data, it may be noted that there is no other phase besides the cubic RS phase related to $Cd_xMg_{1-x}O$. However, in **Fig. 1**b and **2**b, double or triple peaks start to appear. It could be evidence of effective but inhomogeneous incorporation of Mg ions into the CdO lattice. Therefore we observe two or three XRD peaks. Interestingly, on *m*-plane sapphire, this effect was evidenced for the sample with 16, 34, and 42% of Cd, whereas from *c*-plane sapphire this effect was observed for samples with a small concentration of Cd (4% and 16%). The presence of double XRD peaks in $Cd_xMg_{1-x}O$ layers grown by MOCVD was previously reported by Guia *et al.*[23]. It was related to changes in the growth temperature. In our study, we have grown series of CdMgO layers, with a dedicated concentration of Cd, on two differently oriented substrates exactly with the same growth parameters (growth temperature, Cd and Mg fluxes) at the same MBE process. Thus, this is evidence that not only growth temperature influences on quality and homogeneity of layers but also growth direction or orientation of substrate plays an important role. The presence of double diffraction peaks is an evidence of the existence of regions with two different concentrations in the CdMgO ternary layers.

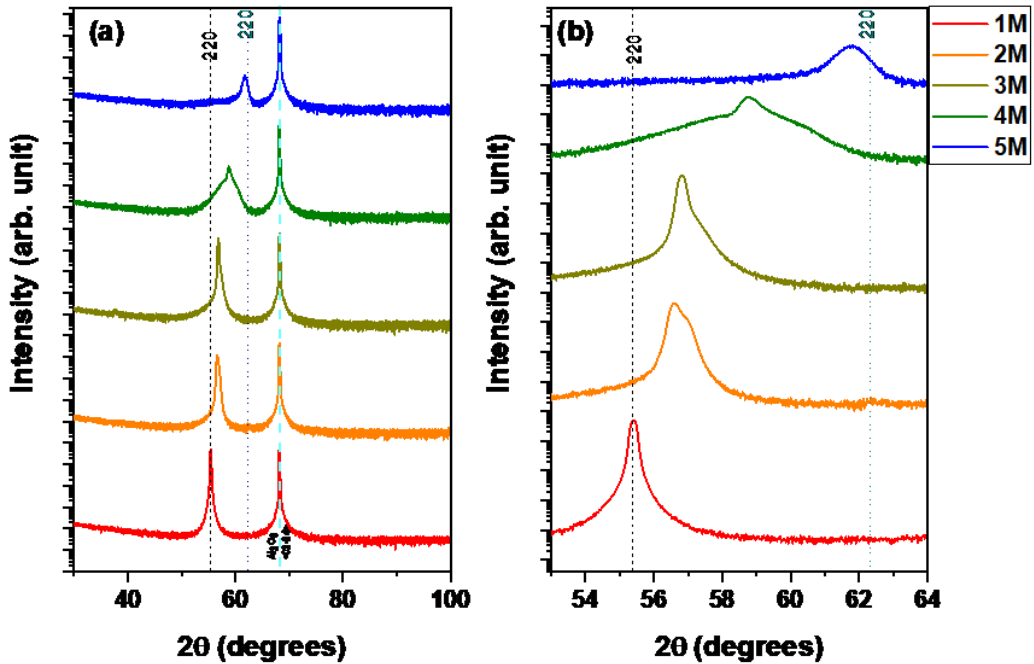

**Fig. 1.** (a) Full X-ray diffraction patterns of series of CdMgO/*m*-$Al_2O_3$ with different Cd concentrations. (b) 220 XRD peak shift.

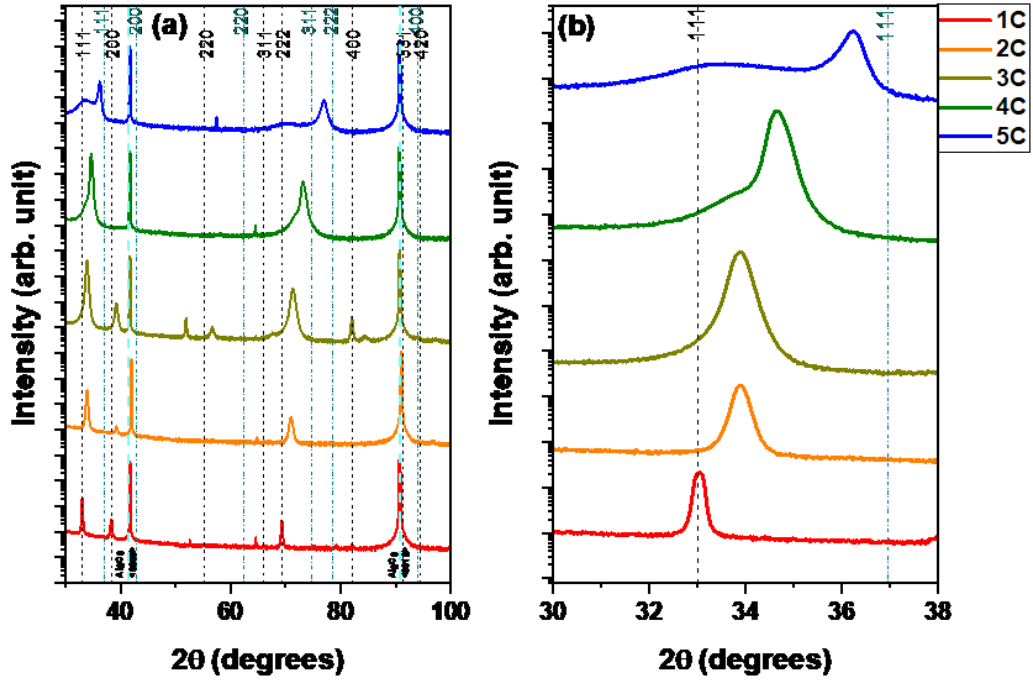

**Fig. 2.** (a) Full X-ray diffraction patterns of series of CdMgO/*c*-Al$_2$O$_3$ with different Cd concentrations. (b) 111 XRD peak shift.

Shifting and broadening of the diffraction peaks were shown in **Fig. 1**b and **Fig. 2**b. Based on this data, the lattice parameter of Cd$_x$Mg$_{1-x}$O cubic alloys can be determined using Bragg's equation:

$$2d\sin\theta = n\lambda \qquad (1)$$

$$a = d * (h^2 + k^2 + l^2)^{\frac{1}{2}} \qquad (2)$$

Where *d* is interplanar distance, $\lambda$ is the X-ray wavelength used, $\theta$ is the Bragg angle, *a* is the lattice parameter, and *(h, k, l)* are Miller indices of Bragg plane. The diffraction peaks were analysed by fitting its Voight function. In order to separate size and micro strain effects, in Voigt peak analysis FWHMs components (Cauchy and Gauss) were determined. The grain size of the alloys was calculated using the Scherrer formula:

$$\tau = \frac{k\lambda}{\beta_c \cos\theta} \qquad (3)$$

where $\tau$ is grain size, *k* is shape factor – Scherrer constant (k= 0.94), $\beta_c$ is Cauchy component of diffraction peak corrected by instrumental resolution [32,33]. Whereas, the micro strain *e* was calculated with the use FWHM Gaussian contribution $\beta_G$. corrected by instrumental resolution using relation [34]

$$e = \beta_G / 4\tan\theta \qquad (4).$$

It was measured that the instrumental errors are equal to $\delta\beta_G$ ±0.0035° and $\delta\beta_C$ ±0.0014°. In ours case, the calculated lattice parameters, grain sizes and micro strain of each sample are listed in **Table 1** and **Table 2** for CdMgO alloys grown on *m*- and *c*-plane sapphire, respectively. It is observed that with an increase in Mg concentration into CdO lattice, the peak broadening occurs which is consistent with a decrease in lattice parameter and alloy grain size as calculated. The most broadened XRD peak is registered for $Cd_{0.04}Mg_{0.96}O$ sample on both the *m*-and *c*-plane sapphire substrates. The increasing of Mg concentration causes the widening and shifting of the (220) (on-*m*-plane $Al_2O_3$) peak. It is characteristic behaviour for alloys and for sample with 4% Cd content peak is located close to the position of pure MgO. The broadening of the 220 and 111 peaks with increasing Cd concentration shown in **Fig. 3** indicates that the grain size of the film decreases with an increase of Mg concentration. Obtained values of the grain sizes are much bigger than those reported for 220 peaks for CdMgO layers obtained by radio frequency magnetron sputtering what indicating better crystal quality of the MBE layers[21]. It has been shown that when the CdMgO layers were grown on *c*-plane sapphire, the Mg concentration has a smaller effect on the grain sizes. The values of micro strain obtained by applied size-strain method increase with Mg content for both series of samples. The increase is more significant for samples grown on *m*-sapphire than for samples on *c*- sapphire. Obtained values of micro strain (table 1 and 2) are comparable with data reported for Mg doped CdO films obtained by spray pyrolysis[35], and for CdO-NiO nanocomposite obtained by chemical precipitation method[36].

| sample | 2Θ (degrees) | $a$ (nm) | $\beta_c$ (degrees) | Grain size (nm) | $\beta_G$ (degrees) | Micro Strain ($*10^{-4}$) |
|---|---|---|---|---|---|---|
| 1M | 55.40 | 0.46845 | 0.023 | 415 | 0.20 | 16.4 |
| 2M | 56.63 | 0.45909 | 0.035 | 269 | 0.24 | 19.2 |
| 3M | 56.82 | 0.45629 | 0.044 | 210 | 0.22 | 17.2 |
| 4M | 58.78 | 0.45768 | 0.418 | 22 | 0.28 | 21.3 |
| 5M | 61.50 | 0.45432 | 0.315 | 29 | 0.65 | 46.9 |

**Table 1** Lattice constant, FWHM (Cauchy-$\beta_c$, Gaussian -$\beta_G$ components) and Grain sizes and micro strains in CdMgO/*m*-$Al_2O_3$

| sample | 2Θ (degrees) | $a$ (nm) | $\beta_c$ (degrees) | Grain size (nm) | $\beta_G$ (degrees) | Micro Strain ($*10^{-4}$) |
|---|---|---|---|---|---|---|
| 1C | 33.03 | 0.46915 | 0.094 | 91 | 0.20 | 28.3 |
| 2C | 33.84 | 0.45824 | 0.133 | 64 | 0.23 | 31.3 |
| 3C | 33.88 | 0.45602 | 0.125 | 68 | 0.26 | 36.8 |
| 4C | 34.66 | 0.45771 | 0.103 | 83 | 0.27 | 36.3 |
| 5C | 36.21 | 0.45758 | 0.161 | 53 | 0.29 | 37.9 |

**Table 2** Lattice constant, FWHM (Cauchy-$\beta_c$, Gaussian -$\beta_G$ components) and Grain sizes and micro strains in CdMgO/*c*-$Al_2O_3$

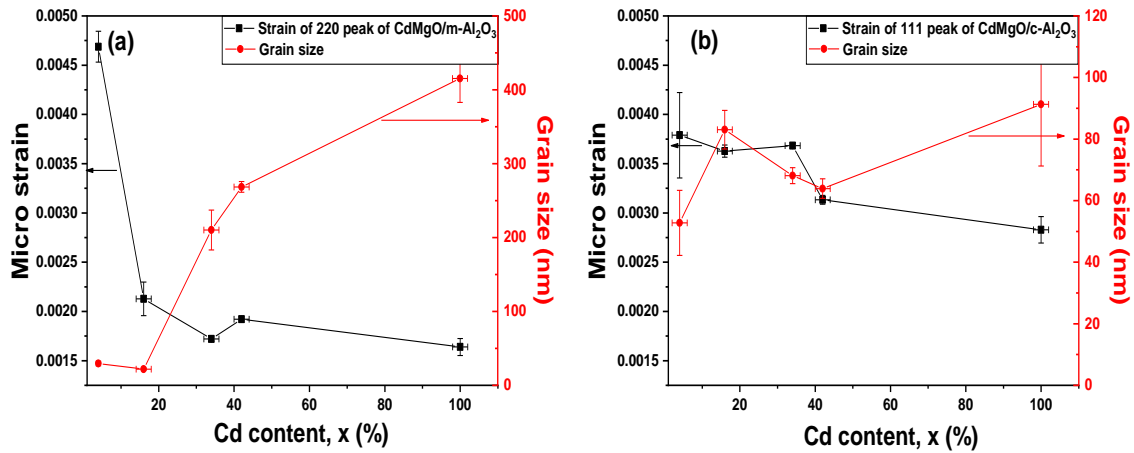

**Fig. 3.** Variation of Micro strain and grain size for CdMgO layers with the increase of Cd concentration in case of (a) 220 peak of CdMgO/*m*-sapphire (b) 111 peak of CdMgO/*c*-sapphire.

### 3.2. EDX study

The series of samples were subjected to EDX analysis to examine Cd and Mg contents in the $Cd_xMg_{1-x}O$ random alloys. In **Fig.4,** the EDX spectra of selected CdMgO samples with different Cd content are shown; O (K), Cd (L) and Mg (K) lines are clearly visible. It is observed that the Mg content increases with the increasing temperature of the Mg effusion cell in the random alloy. Cd content decreases as the Mg increases, which very well supports the substitution of $Cd^{2+}$ ions with $Mg^{2+}$ ions in the host lattice. The obtained values of individual Cd and Mg contents in atomic percent are listed in **Table 3**. The measurements error of Cd and Mg contents was on the level of ± 2 at.%.

| Sample | Cd% | Mg% | Mg effusion cell temp (°C) | Flux$_{Cd}$ (Torr) | Flux$_{Mg}$ (Torr) |
|---|---|---|---|---|---|
| **1** | 100 | - | - | 1,25e-7 | --- |
| **2** | 42 | 58 | 505 | 1,21e-7 | 0,10e-7 |
| **3** | 34 | 66 | 520 | 1,37e-7 | 0,30e-7 |
| **4** | 16 | 84 | 530 | 0,98e-7 | 0,22e-7 |
| **5** | 4 | 96 | 545 | 0,87e-7 | 0,43e-7 |

**Table 3** Results of EDX analysis in CdMgO random alloys and Cd and Mg fluxes

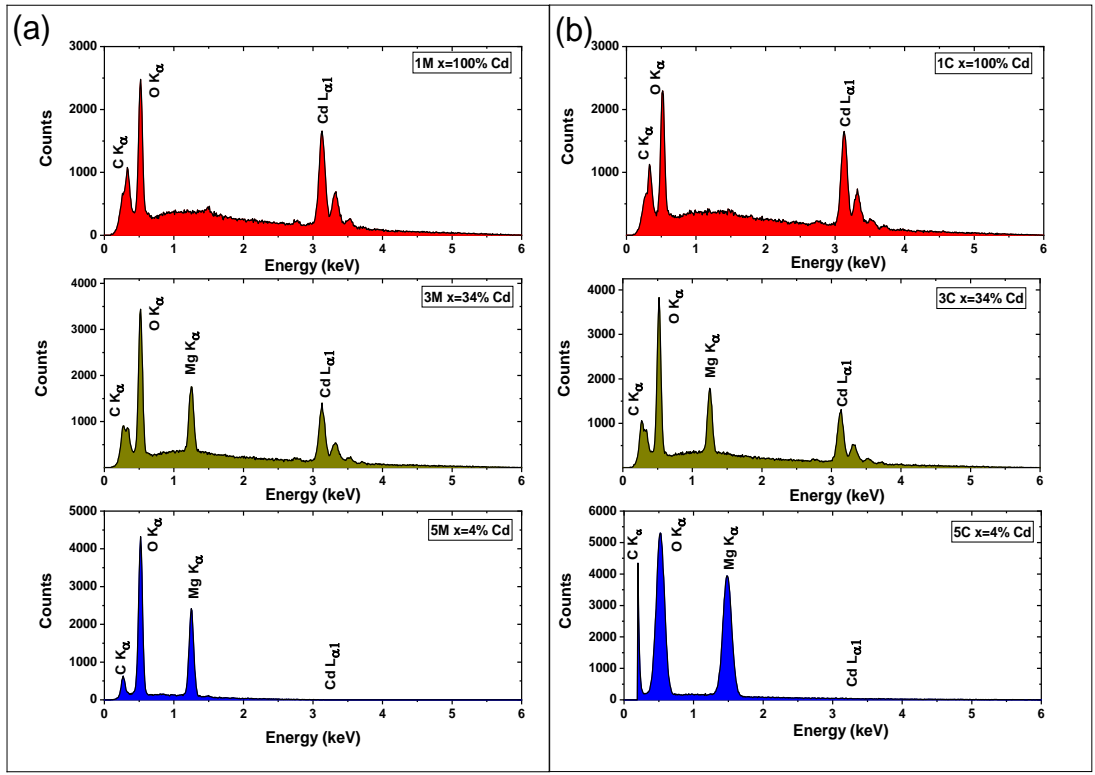

**Fig. 4**. EDX spectra of selected CdMgO layers grown on (a) *m*-plane and (b) on *c*-plane sapphire substrates.

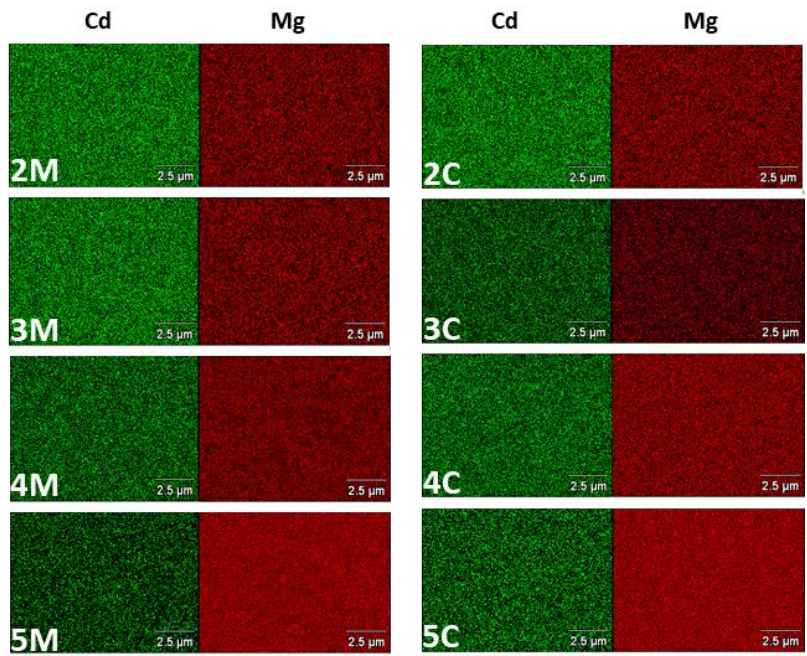

**Fig. 5**. Elemental distribution maps of Cd and Mg in CdMgO alloys grown on *m*- and *c*-plane sapphire substrates. Green and red colour represents Cd and Mg elements respectively.

EDX has been carried out also in the mode of elemental mapping and was presented in **Fig. 5**. EDX maps gave us information about the uniform distribution of Cd and Mg elements in CdMgO alloys. Previous study by Guia *et al.* [22,23] reveals, formation of regions with different concentrations of Cd in CdMgO alloys prepared at higher growth temperatures. However, in our case, we do not observe this effect.

The dependence of the lattice parameters of CdMgO random alloys with Cd content (appointed with EDX study) is shown in **Fig. 6**. To calculate the lattice parameters of $Cd_xMg_{1-x}O$ alloys, the lattice parameter of 0.46953 nm and 0.4213 nm for pure CdO and MgO are used, respectively[5]. All diffraction peaks are fitted with Voight function and the lattice parameters were calculated. It is observed that the obtained lattice parameters deviate from Vegard's law which is

$$a_{CdxMg1-xO} = xa_{CdO} + (1-x)a_{MgO}. \quad (5)$$

Hence we have fitted our data to a parabolic function using the modified Vegard's law:

$$a_{CdxMg1-xO} = xa_{CdO} + (1-x)a_{MgO} - bx(1-x) \quad (6)$$

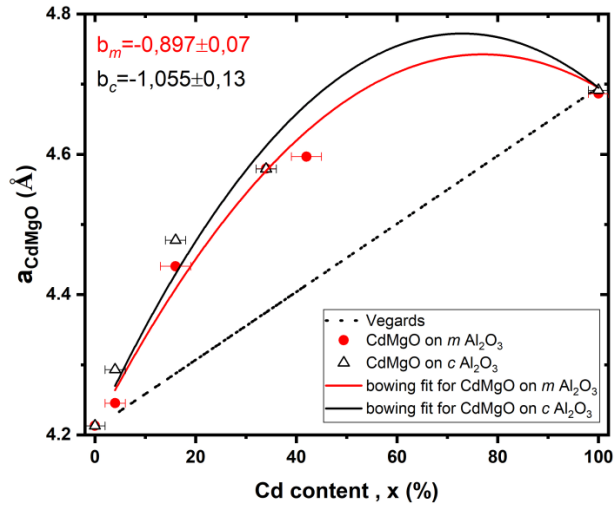

**Fig. 6.** Lattice parameters of the CdMgO alloy vs Cd concentration.

where $a_{MgO}$, $a_{CdO}$, $a_{CdxMg1-xO}$ are lattice parameters of MgO, CdO, and CdMgO, respectively. Next, x is the atomic concentration of Cd obtained from the EDX study and b is the bowing parameter. From fitting the lattice parameters of CdMgO alloys, the bowing constant was found to be -0.897 nm and -1.055 nm for CdMgO/*m*-$Al_2O_3$ and CdMgO/*c*-$Al_2O_3$, respectively. Earlier theoretical calculation of bowing parameter using density functional theory found that it is -0.214 nm[37]. Guia *et al.* studied MgCdO thin film on a *c*-plane sapphire substrate and reported the bowing constant to be -0.449 nm[23]. The negative bowing represents upward bowing and it suggests repulsive interaction between O-2p and Cd-3d states[38].

### 3.3. UV –Vis spectroscopy

The optical properties of CdMgO random alloys on $m$ and $c$– plane sapphire substrates were investigated at room temperature using a UV- Visible spectrophotometer in the wavelength range of 200-600 nm. The transmittance spectra are plotted as a function of wavelength in **Fig. 7**a and **Fig. 8**a for CdMgO samples on $m$-plane and $c$-plane sapphire, respectively. From the experimental data, the spectral absorption coefficient $\alpha(\lambda)$ is calculated using the relation $\alpha = (1/d)\ln(1/(\%T))$, and the optical bandgap calculated using the Tauc relation,

$$\alpha h\nu = A\,(h\nu - E_{g,o})^k \qquad (7)$$

Where d is the film thickness, $h\nu$ is the photon energy in eV, $A$ is a constant, $E_{g,o}$ is the optical bandgap, and $k$ is a constant which depends upon the type of electronic transition. The value of $k$ is equal to ½ and 2 for direct transition and indirect transition respectively[39]. The direct optical bandgap can be calculated by extrapolating the linear portion of $\alpha^2$ to the energy axis as shown in **Fig. 7**b and **Fig. 8**b for CdMgO random alloys on $m$-plane and $c$-plane sapphire substrate respectively. As we know for CdO, apart from the direct bandgap, it has two indirect bandgaps[8,9,29]. Hence we have calculated the indirect bandgap for samples having higher Cd concentration (i.e. samples 1, 2, and 3) by extrapolating the linear region of $\alpha^{1/2}$ to the energy axis as shown in **Fig 7**c and **8**c.

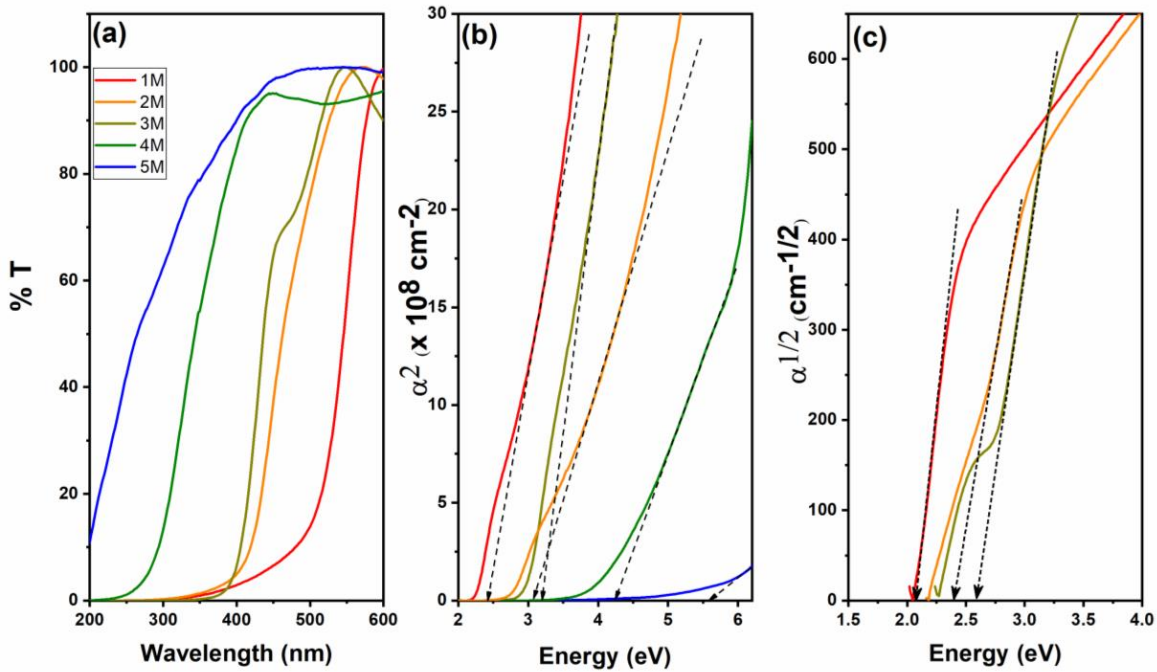

**Fig. 7**. (a) Transmittance spectra of CdMgO random alloys, (b) $\alpha^2$ plot as a function of photon energy (hν), and (c) $\alpha^{1/2}$ plot as a function of photon energy (hν) for CdMgO films on $m$-sapphire.

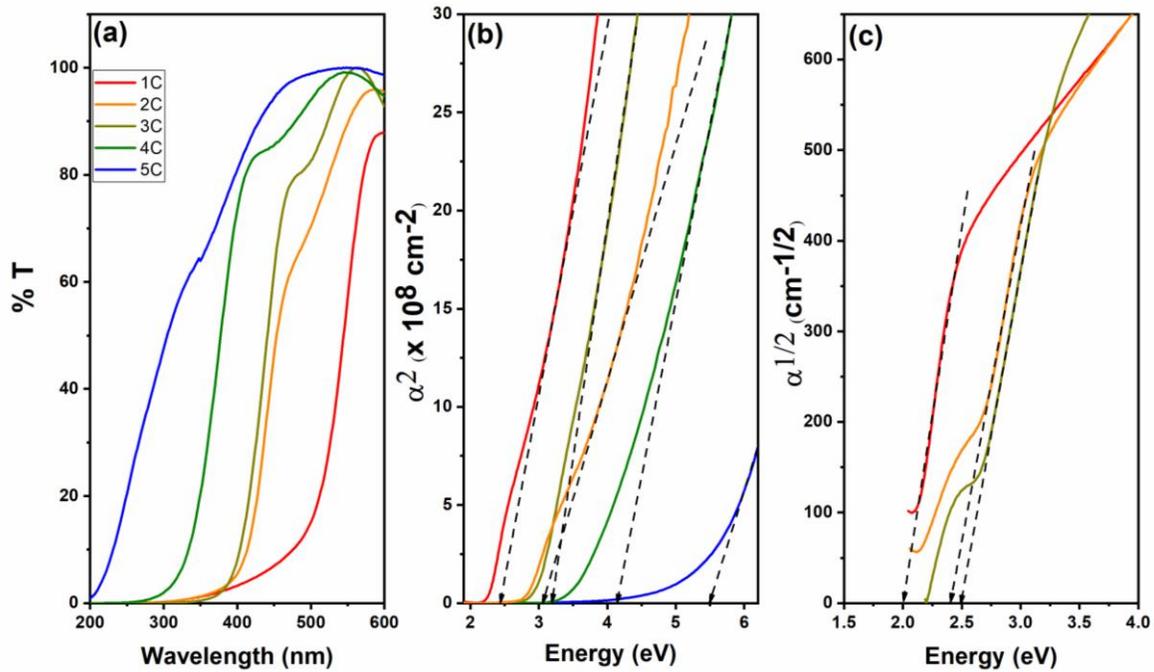

**Fig. 8.** (a) Transmittance spectra of CdMgO random alloys, (b) $\alpha^2$ plot as a function of photon energy (hv), and (c) $\alpha^{1/2}$ plot as a function of photon energy (hv) for CdMgO films on *c*-sapphire.

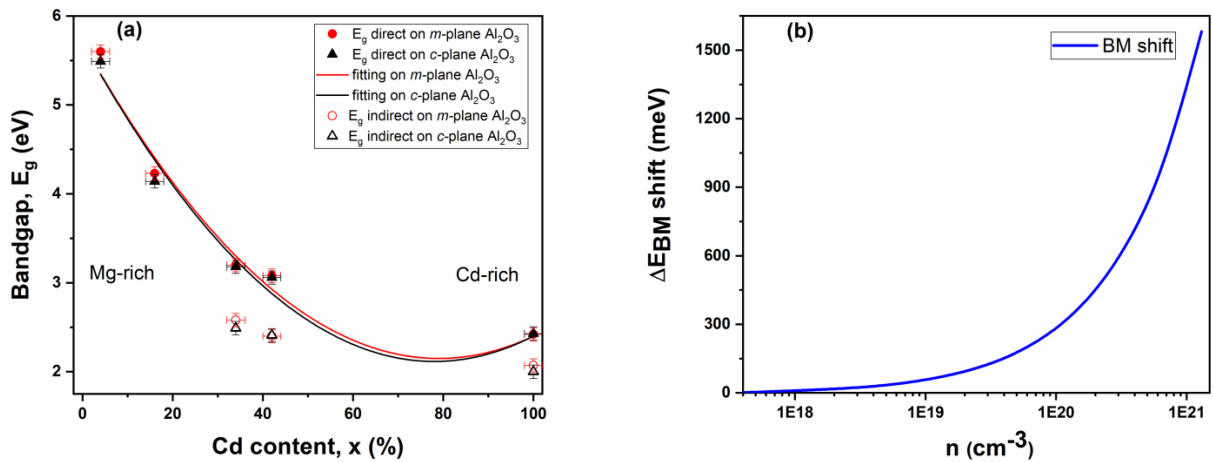

**Fig. 9.** (a) Variation of optical direct bandgap (full symbols) and indirect (open symbols) with an increase in Cd concentration in CdMgO alloy on *m*- and *c*-plane sapphire (red and black points indicate for CdMgO/*m*-Al$_2$O$_3$ and CdMgO/*c*-Al$_2$O$_3$, respectively) (b) calculation of BM energy shift as a function of carrier concentration for CdO.

The high transparency of films in the visible regime suggests that CdMgO alloys are good crystalline and can be used for TCOs application purposes. For pure CdO the direct bandgap is found to be 2.42 eV. The optical bandgap is blue-shifted with increases in Mg concentration as shown in **Fig. 9**a. The obtained optical direct bandgaps are fitted using the modified Vegard's law

$$(E_g)_{Cd_xMg_{(1-x)}O} = x(E_g)_{CdO} + (1-x)(E_g)_{MgO} - bx(1-x) \qquad (8)$$

where $Eg_{CdO}$ and $Eg_{MgO}$ are optical bandgaps of CdO and MgO in the rocksalt crystal structure. The value of $Eg_{CdO}$ and $Eg_{MgO}$ is equal to 2.4 and 5.7 eV, respectively. Hence, we fit the experimental optical bandgap data using equation (7) as shown in **Fig. 9**a. The best fit was obtained for a bowing parameter value of 5.68 ± 0.49 and 5.88± 0.48 eV for CdMgO alloy on *m*-plane and *c*-plane sapphire, respectively. Earlier reported bowing values are 2.17eV and 3.1eV for CdMgO alloys on glass and sapphire substrates[19, 22]. The Mott criterion of pure CdO is fulfilled for the carrier density of $N_c = 3 \times 10^{18}$ cm$^{-3}$[40], subsequently for higher carrier concentration CdO should be regarded as a degenerate semiconductor. Using recent literature[41] we have found that for the CdO sample the bandgap widening can occur due to Burstein–Moss effect and it is significant for carrier concentration n> $10^{19}$ cm$^{-3}$ (as shown in **Fig. 9**b). The B-M shift ($\Delta E_{BM}$) can be expressed using, $\Delta E_{BM} = \frac{\hbar^2(3\pi^2 n)^{2/3}}{2m^*}$, where n is the carrier concentration, $m^*$ is the reduced effective mass of CdO (for CdO, $m^* = 0.17\ m_0$)[42]. However, the incorporation of Mg ion into CdO lattice results in a decrease in electron concentration in the alloy[21]. Importantly, we analyzed samples with a wide range of Cd concentration (*x* values in Cd$_x$Mg$_{1-x}$O alloys changes from 100 to 4%). Due to Mg doping in ternary alloys concentration of electrons rapidly decreases[21], so in the analyzed set of samples we can consider the strong influence of the B-M only for pure CdO.

### 3.4. AFM study

**Fig.10** shows the surface morphology measured by Atomic Force Microscopy (AFM) of the CdMgO random alloys on *m*-plane and *c*-plane sapphire substrates over 5 μm x 5 μm scan area. As can be seen, a variation in morphological aspect is observed with a change in Cd content. The obtained average roughness ($R_a$) values are summarized in **Table 4** for CdMgO alloys on *m*- and, *c*-plane sapphire. As shown in **Fig. 11**, for CdMgO/*m*-Al$_2$O$_3$ series of samples, 3M (34% Cd content) alloy films have the minimum $R_a$ value. Whereas the minimum roughness parameters have been observed in the alloy film having 42% Cd content (2C), in the case of CdMgO/*c*-Al$_2$O$_3$ series of sample. The minimum roughness parameters indicate that these alloys have fewer grain cavities and better surface uniformity compared to other alloys. However, the surface roughness value is higher for Cd-rich conditions. The number of papers about AFM study in case of CdMgO oxide is very limited. The changes in the average roughness parameter ($R_a$) are depending on the structure and growth parameters e.g.: growth temperature [43]. The reason for the morphological change in our samples is the change in the alloy concentration. The growth variables such as oxygen plasma parameters (oxygen gas flow, RF power) were the same for all investigated layers. It is well known that growth parameters for pure CdO and MgO oxides usually differ. Thus changing the concentration in CdMgO alloy can implicit changes in rms parameter. It was presented that CdO:Mg films obtained by spray pyrolysis on glass substrates (with Mg concentration 15% and 30%) show a smooth surface compared to the undoped CdO films[44]. Mentioned results for relatively low Mg content in CdMgO are quite consistent with our data (samples 2 and 3). In our case whole Mg concentration region were investigated. The variation of surface morphology reveals the importance of different growth conditions as well as different oriented sapphire substrates on which CdMgO alloys are grown.

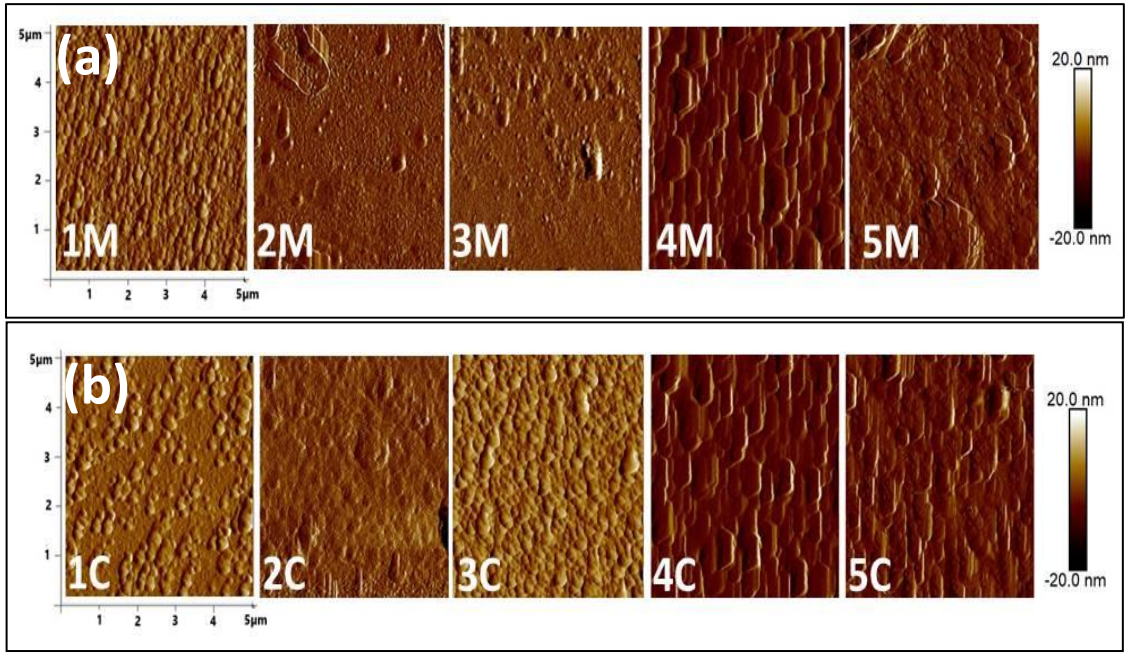

**Fig. 10.** AFM images of CdMgO alloys on (a) *m*-sapphire, and (b) *c*-plane sapphire.

| Sample | Cd% (from EDX) | $R_a$(nm) *m*-plane | $R_a$(nm) *c*-plane |
|---|---|---|---|
| 1 | 100 | 6.2 | 8.0 |
| 2 | 42 | 3.2 | 1.6 |
| 3 | 34 | 1.6 | 10.9 |
| 4 | 16 | 9.6 | 10.4 |
| 5 | 4 | 12.9 | 7.8 |

**Table 4** Roughness Values of CdMgO on *m*- and *c*-$Al_2O_3$

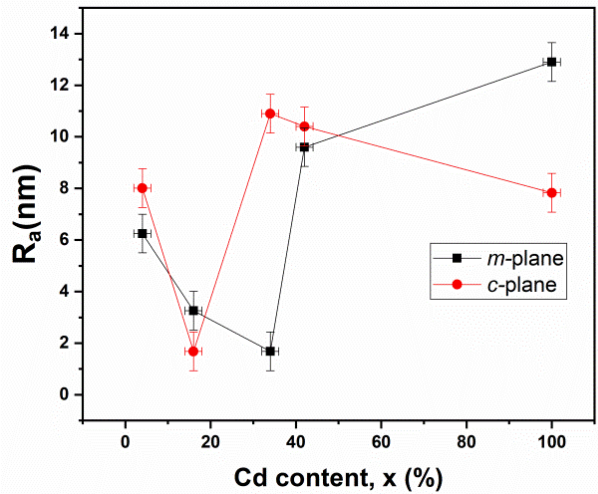

**Fig. 11.** Average roughness value with increase in Cd concentration in CdMgO alloys.

## 4. Conclusions

In this work, we have grown CdMgO series of alloys on *m*- and *c*-plane of sapphire substrates using plasma-assisted molecular beam epitaxy technique. The change in growth temperature of Cd and Mg effusion cells results in Cd-rich to Mg-rich growth conditions. Moreover, a significant change in structural and optical properties has been observed in between these two growth conditions. Structural, optical, and morphological characterizations were done using XRD, EDX, UV-Vis spectroscopy, and AFM.

From the XRD study, we found that there is a significant decrease in grain size with an increase in Mg substitution in case of CdMgO alloy on *m*-plane sapphire substrate whereas, the grain size does not vary drastically in case of CdMgO alloy on *c*-plane sapphire. The lattice constants also decrease with an increase in Mg content in the alloys. The lattice parameters were fitted using modified Vegard's law and the bowing constant was found to be -0.0826 and -0.1003 nm for CdMgO/*m*-$Al_2O_3$ and CdMgO/*c*-$Al_2O_3$ respectively. Based on XRD results it can be concluded that CdMgO layers on *m*-plane sapphire substrates have better crystallographic quality than those on *c*-plane sapphire substrates.

In analyzed layers, the direct bandgap was varied from 2.42 eV to 5.5 eV by changing Mg concentration from 0 to 96%. The bandgap values depend nonlinearly with Mg concentration and the bowing parameter found to be 5.68 ± 0.49 and 5.88 ± 0.48 eV for CdMgO alloy on *m*-plane and *c*-plane sapphire respectively. The indirect bandgap is obtained for samples with higher Cd content (*i.e. samples 1, 2, and 3*). The indirect bandgap changes from 2.07 to 2.58 eV and 2 to 2.49 eV for CdMgO on *m*-plane and *c*-plane sapphire respectively with an increase in Mg concentration from 0 to 66%.

EDX mapping did not reveal the inhomogeneous distribution of Cd and Mg atoms in the alloys in contrast to previously reported data[23]. Morphological study using AFM demonstrated that surface roughness of CdMgO alloys depends on Mg concentration. For the first time, we have reported the roughness parameter value for the CdMgO series of samples grown on *c*- and

*m*-oriented sapphires. In general, for these series of samples, the lowest RMS values were obtained for samples with a high Mg content.

Due to wide bandgap changes with Mg incorporation, we expect this work will shed light on Cd-rich to Mg-rich CdMgO alloys and opens up new prospects for the design of novel optoelectronic devices.

**Acknowledgment**

This research was funded in whole by the National Science Center, Poland, Grant No. 2021/41/N/ST5/00812, and 2021/41/B/ST5/00216. For the purpose of Open Access, the author has applied a CC-BY public copyright license to any Author Accepted Manuscript (AAM) version arising from this submission.